\documentstyle[palatino,psfig,11pt]{article}
\begin{document}

\begin{center}
{\Large Nuclear interference effects in $^8$B sub-Coulomb breakup} \\
\vspace{0.2cm}

F.M. Nunes\footnote{Email address: filomena@wotan.ist.utl.pt} \\
{Departamento de F\'{\i}sica, CENTRA, Instituto Superior T\'ecnico, \\
Av. Rovisco Pais, 1096 Lisboa, Portugal.}  \\
\vspace{0.2cm}

I. J. Thompson\\
{Department of Physics, University of Surrey, Guildford GU2 5XH, UK}  
\end{center}

\begin{abstract}
The breakup of $^8$B on $^{58}$Ni below the Coulomb barrier was measured
recently with the aim of determining the Coulomb breakup components. 
We reexamine 
this reaction, and perform one step quantum-mechanical  calculations that 
include
E1, E2 and nuclear contributions. We show that the nuclear contribution 
is by no means negligible at the intermediate angular range where data was 
taken.
Our results indicate that, for an accurate description of this reaction, 
Coulomb E1, E2 and nuclear processes all have to be taken into account.
\end{abstract}

In order to understand the details of Coulomb breakup experiments of
$^8$B on heavy targets \cite{moto1,moto2}, and to extract an $S_{17}$ 
for astrophysical relative $p+^7$Be energies
from the measured Coulomb Dissociation (CD) cross sections, 
 it is necessary to know the
relative importance of the E1 and E2 Coulomb contributions to breakup.
The  Notre Dame experiment  on the Coulomb dissociation of $^8$B on
$^{58}$Ni at 26 MeV \cite{nd} therefore sought to determine these
contributions by measuring the integrated cross section of the $^7$Be
fragment between 39$^\circ$ and 51$^\circ$. At these much lower
beam energies the E2 should be as large (or larger) than the E1
contribution, and so a measurement of the total breakup probability
should give strong constraints on both the integrated B(E1) and B(E2)
transition strengths. Since the closest distance of approach at 26 MeV
for the above angles is 15 fm, and plausible optical potentials give
elastic $\sigma/\sigma_R< 0.5$ only beyond 90$^\circ$, it was believed
that the E1 and E2 contributions could be measured free from nuclear
effects with the Notre Dame setup.

However, the results of the Notre Dame experiment \cite{nd}  disagreed
\cite{shyam,puri} with the theoretical predictions for a variety of
structure models of $^8$B (\cite{kim,esb}), when using the standard
semiclassical theory of Coulomb excitation \cite{alder}.  The
predictions of the semiclassical breakup theory were twice or three
times the measured cross section.  This discrepancy raises some
fundamental questions: How strong is the dependence of the CD cross
section on structure model of $^8$B?  Should nuclear effects be taken
into account? How important are interference terms: Coulomb-Coulomb,
nuclear-nuclear and Coulomb-nuclear?  In ref. \cite{puri} it is shown
that no remotely-reasonable structure model could give a breakup
probability as small as that measured in \cite{nd}.  This work answers
the second question on nuclear contributions,  and  hints on a possible
answer to the third question on multistep effects.

In order to examine the validity of the semiclassical approximations
previously used in \cite{nd,shyam}, we performed quantum-mechanical
calculations of the single-proton excitations from a ground state to a
range of continuum states.  We discretise the continuum by the method
of continuum bins, as used in the CDCC methods reviewed in ref.
\cite{cdcc}. Since, however, results using this method for
Coulomb transitions have been attempted \cite{saku}, but not yet fully
demonstrated \cite{rebel} to be converged, especially for E1 transitions, we confine
ourselves to dipole and quadrupole transitions only to and from the
ground state to the continuum, and omit the couplings between continuum states. We
hence perform first-order distorted-wave Born approximations to the
full CDCC problem, but report on these DWBA results because these
immediately lead us to see some severe short-comings of the previous
applications of semiclassical methods.

We therefore perform prior-DWBA breakup calculations by discretising
the continuum with $s_{1/2}$, $p_{1/2}$, $p_{3/2}$, $d_{3/2}$,
$d_{5/2}$ partial waves up to E(p-$^7$Be)$=3$ MeV.   We have taken a
single particle model for the structure of $^8$B assuming that all
states in $^8$B are determined by the g.s. potential defined in
\cite{esb}  (this simplification of the $^7$Be-p scattering state
interaction  has hardly any effect on the integrated Coulomb Dissociation cross
section \cite{puri}).  For the continuum discretisation, good accuracy
is obtained if we use 13 bins defined in the following way: $9$ bins of
$100$ keV centred at $0.15; 0.25; ...; 0.95$ MeV and $4$ bins of $500$
keV centred at 1.25; 1.75; 2.25; 2.75 MeV. In order to obtain
convergence for the E1 and E2 transitions we include up to $l_{max}=600
\hbar$ and $R_{max}=300$ fm for the reaction mechanism, in the code {\sc
Fresco} \cite{fresco}.

\begin{figure}[h]
\centerline{
	\parbox[t]{0.8\textwidth}{
\centerline{\psfig{figure=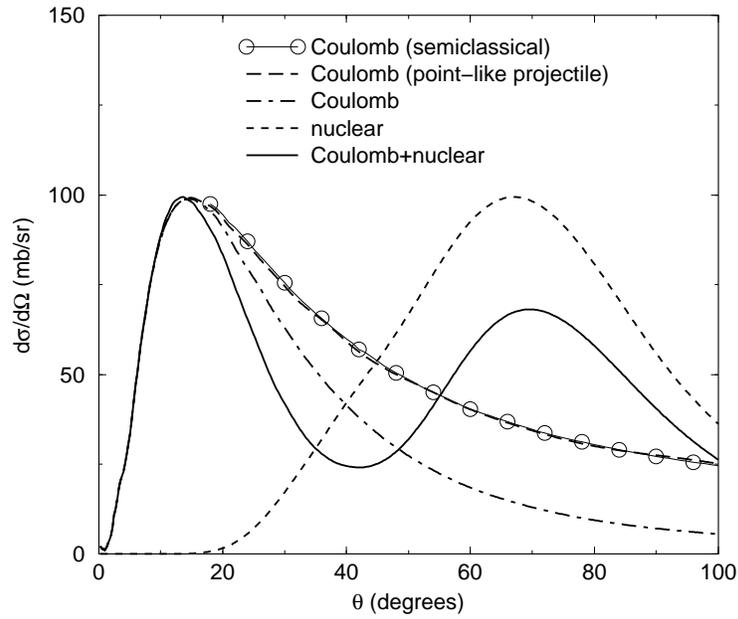,width=0.6\textwidth}}
	\caption{Comparison of the Coulomb and the nuclear contributions to the 
differential cross section for the breakup of $^8$B on $^{58}$Ni
in the Notre Dame experiment \protect\cite{nd}.}	
\label{fig:1}}
}
\end{figure}

We first check that we reproduce the pure Coulomb semiclassical results
for E1 and E2 excitations under the same physical approximations.  This
requires that we use the pure $r^{- \lambda -1}$ shapes for the Coulomb
multipoles, which would be true for a point projectile.  The comparison
of the semiclassical and quantum mechanical differential cross section
as a function of the $^8$B scattering angle is shown by the circles and
long-dashed lines in fig.(\ref{fig:1}) and the agreement is perfect.
This indicates that the continuum energy range has been discretised
with sufficient accuracy for at least the one-step treatment of this
reaction.

In fig.(\ref{fig:1}) the dot-dashed line shows
the pure Coulomb result obtained by
folding the projectile-target interaction with a $^8$B wavefunction
of a realistic size (the $^7$Be-p interaction given by \cite{esb}).
Due to the long tail in the $^8$B g.s. wavefunction (the binding energy
is only $0.137$ MeV)  we find that the point-projectile approximation
is not valid for angles larger than $20 ^{\circ}$. The simple condition
$b > R_p + R_T$ assumed in applying the semiclassical Alder and Winther theory
is not valid when the projectile has such an extended nature.

We next calculate the pure nuclear differential cross section
(short-dashed line in fig. \ref{fig:1}).  We use a Becchetti-Greenlees
p-$^{58}$Ni potential \cite{ becc} at  3 MeV,
and we take, for
the $^7$Be-$^{58}$Ni,  the optical  potential from \cite{glov} which was
extracted from $^7$Li scattering on $^{58}$Ni at $34$ MeV.  
These potentials, with their Coulomb parts,
are both used in folding integrals to find dipole
and quadrupole transitions from the ground state. They are also
folded to obtain the rather diffuse monopole potentials which
govern the c.m. motion of the excited $^8$B$^*$ states in the 
exit channels. We find (fig. \ref{fig:1}) 
that the nuclear contribution is insignificant up to $20 ^{\circ}$, but
grows rapidly beyond that,  peaking at $\simeq 70 ^{\circ}$.

When nuclear and Coulomb multipoles are included coherently (solid line
in fig.\ref{fig:1}), there are already small effects below
$20^{\circ}$, a pronounced Coulomb-nuclear interference minimum between
$25^{\circ}$ and $50 ^{\circ}$, and a nuclear-dominated peak at $\simeq
70^{\circ}$.  This large nuclear effect is present even though the
elastic Coulomb + nuclear cross section  drops only to $90\%$ of the
Rutherford cross section at $70 ^{\circ}$, because of the large
halo-like size of proton wavefunction in the g.s. of $^8$B.  The dip in
the differential cross section coincides with the angular range
measured in the Notre Dame experiment, suggesting that including
nuclear effects is at least part of the solution to the disagreement
between semiclassical predictions and data.  Our calculations of the
nuclear effects are qualitatively similar to the results in
\cite{vitturi},  where Coulomb and nuclear effects are also calculated
by folding single-particle potentials over the wave functions of
discretised continuum states. In \cite{vitturi}, however, the $^7$Be$ +
^{58}$Ni potential is omitted from the transition operator, and the
excitation mechanisms are determined by integrating along a
semiclassical trajectory determined by a fixed $^8$B + $^{58}$Ni
optical potential.

\begin{figure}[h]
\centerline{
	\parbox[t]{0.8\textwidth}{
\centerline{\psfig{figure=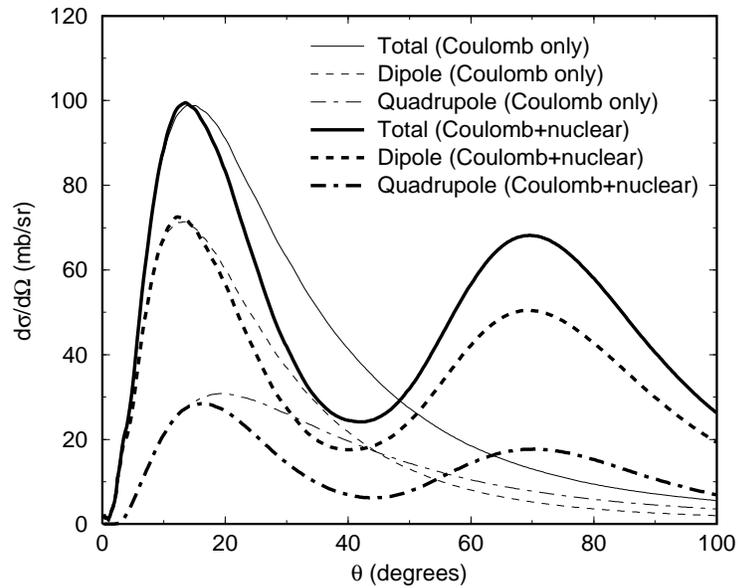,width=0.6\textwidth}}
	\caption{The dipole and quadrupole components of the differential cross 
section for the breakup of $^8$B on $^{58}$Ni with and without the nuclear 
interactions with the target.}	
\label{fig:2}}
}
\end{figure}

In fig.(\ref{fig:2}) we show the separate dipole and quadrupole
contributions to the differential cross section for the pure Coulomb
process (with finite $^8$B size) as well as for the case when nuclear is also 
included.  The
dipole to quadrupole ratio around $45 ^{\circ}$ (the angular range
corresponding to the Notre Dame data)  can change considerably by
including the nuclear contribution. The quadrupole response
is proportionately more affected by the nuclear interference
in the middle range of angles $15^\circ - 50^\circ$.

\begin{figure}[h]
\centerline{
	\parbox[t]{0.8\textwidth}{
\centerline{\psfig{figure=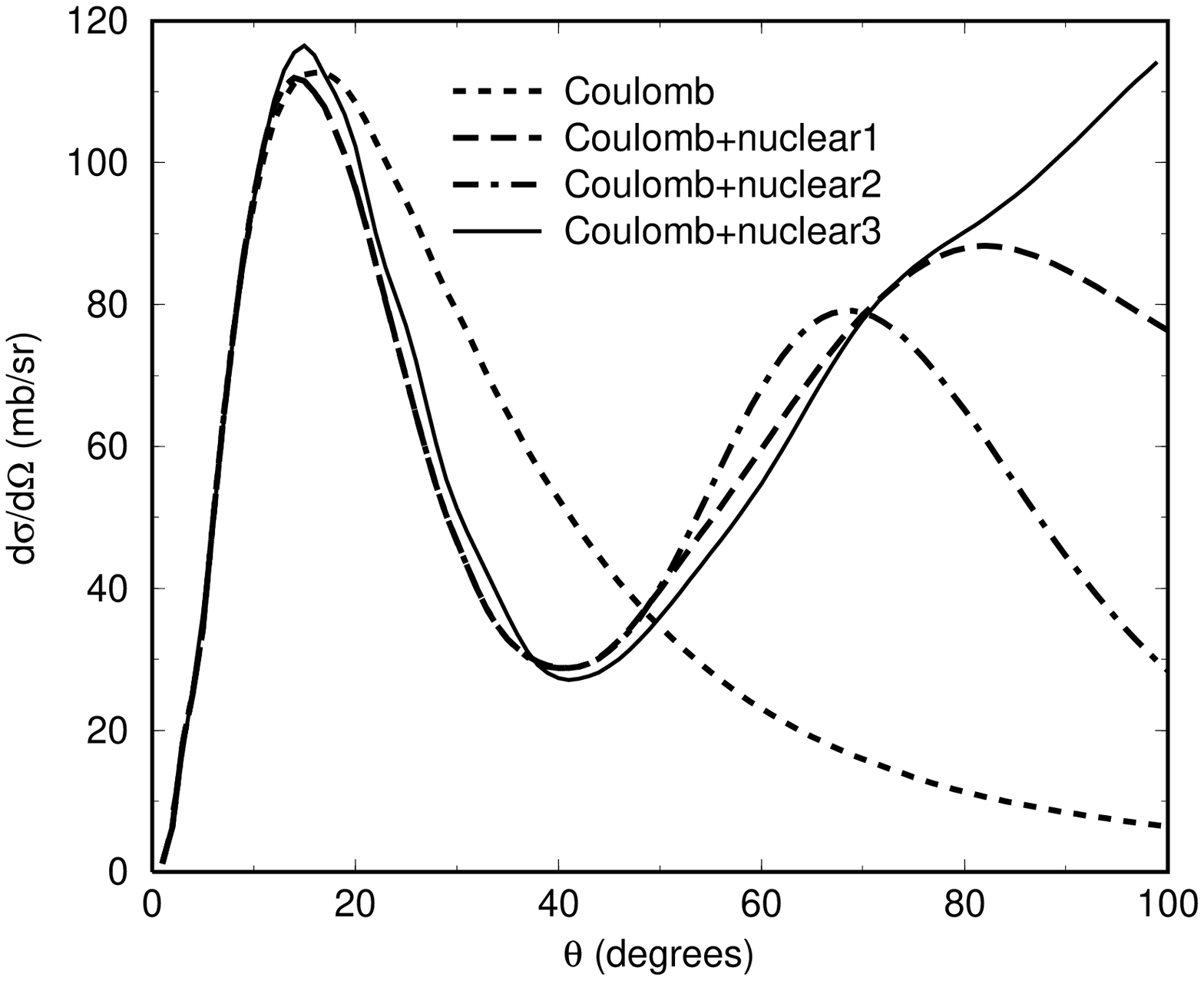,width=0.6\textwidth}}
	\caption{The sensitivity of the differential $^8$B$^*$ breakup cross 
section 
to the $^7$Be-target nuclear interaction.}	
\label{fig:3}}
}
\end{figure}

Finally, in order to illustrate the sensitivity of the breakup cross
section to the optical potentials, we compare in fig.(\ref{fig:3}) the
results using two different sets of parameters both extracted from
$^7$Li scattering data: {\sf nuclear1} from \cite{glov} at $34$ MeV and
{\sf nuclear2} from \cite{moroz} at $14.2$ MeV.  For these calculations we
have also included the relative $f$-waves in the $^8$B continuum, which
increases the cross section because of additional E2 contributions.  Up to $50
^{\circ}$ the differential cross section is unaffected, while there are
large differences around the nuclear peak. We also show the effect of
including only the real part of the $p+^{58}$Ni potential and neglecting the
$^7$Be$+^{58}$Ni potential altogether ({\sf nuclear3}). We see
that the $p+^{58}$Ni potential is responsible for the largest part of the 
interference
effects between $15^{\circ}$ and $50^{\circ}$. Fortunately, this potential is 
experimentally
well-determined as compared with the  $^7$Be potential.
 %Until  the $^7$Be scattering
 %at the relevant energies is measured, the nuclear potentials will be a
 %major source of uncertainty to the extraction of information on the E1
 %and E2 components from low-energy breakup experiments.

 Other authors have pointed out the importance of higher order
Coulomb-Coulomb effects for the $^8$B breakup in the intermediate
energy regime \cite{esb}. One should keep in mind that our conclusions
are based on 1-step distorted-wave coupled continuum bins
calculations.  It is clear from fig.(\ref{fig:1}) that Coulomb-nuclear
interference becomes considerable for angles above $20 ^{\circ}$. Given
the strength of the nuclear peak we would expect multiple step processes to
play an important role.  More work on these lines is underway and will
be reported in the near future.

The initial objective of the Notre Dame experiment was to measure the
magnitude of the E2 component in the CD of $^8$B in order to determine
the smaller E2 effects in the forward angle experiments performed at
higher energy, where the process is E1 dominated \cite{moto1,moto2}.
This low-energy effort becomes increasingly hard when nuclear effects
and multistep processes are mixed in.  
For the unambiguous extraction of E1 and E2 Coulomb amplitudes from
breakup experiments, the large extent of the $^8$B g.s. wave function
requires measurements at 3 MeV/A to be performed at angles
of 15$^\circ$ or more forward. This corresponds to a distance of
closest approach of 40 fm or more. For closer collisions, we find
that nuclear effects cannot be avoided.
Furthermore, more conclusive  results from the Notre Dame experiment
will only be possible after multistep processes have been fully analysed.

\vspace{1cm}
We thank Andrea Vitturi for fruitful discussions.
UK support from the EPSRC grants GR/J/95867
and Portuguese support from JNICT PRAXIS/PCEX/P/FIS/4/96 are acknowledged.
One of the authors, F. Nunes, was supported by JNICT BIC 1481.



\begin{thebibliography}{b8cd}
\itemsep=-4pt
\bibitem{moto1} T. Motobayashi et al., Phys. Rev. Letts. {\bf 73} (1994) 2680
\bibitem{moto2} T. Kikuchi et al., Phys. Letts. {\bf B391} (1997) 261
\bibitem{nd} Johannes von Schwarzenberg et al., Phys. Rev. {\bf C53} (1996) 
R2598
\bibitem{shyam} R. Shyam and I.J. Thompson, Phys. Letts. {\bf B415} (1997) 315
\bibitem{puri} F. M. Nunes, R. Shyam, I.J. Thompson, Proc. {\it International
Workshop on Physics with Radioactive Nuclear Beams}, Puri, India, January 12 - 
17, 1998, submitted to J. Phys. G.
\bibitem{kim} K.H. Kim, M.H. Park and B.T. Kim, Phys. Rev. {\bf C53} (1987) 363
\bibitem{esb} H. Esbensen and G. Bertsch, Nucl. Phys. {\bf  A600} (1996) 37
\bibitem{alder} K. Alder and A. Winther, {\em Electromagnetic Excitation},
(North Holland, Amsterdam, 1975)
\bibitem{cdcc} Y. Sakuragi, M. Yahiro and M. Kamimura, Prog. Theo. Phys. Suppl.
{\bf 89} (1986) 136
\bibitem{saku} Y. Hirabayashi and Y. Sakuragi, Phys. Rev. Lett. {\bf 69} (1992) 
1892
\bibitem{rebel} G. Baur and  H. Rebel, J. Phys. G {\bf 20} (1994) 1
\bibitem{fresco} I.J. Thompson, Computer Physics Reports, {\bf 7} (1988) 167  
\bibitem{becc} F.D. Becchetti and G.W. Greenlees, Phys. Rev. {\bf 182} (1969) 
1190
\bibitem{glov} C.W. Glover, R.I. Cutler and K.W. Kemper, Nucl. Phys. {\bf A 
341} (1980) 137
\bibitem{vitturi} C.H. Dasso, S.M. Lenzi and A. Vitturi, submitted to Nucl. 
Phys. A
\bibitem{moroz} Z. Moroz et al., Nucl. Phys. {\bf A 381} (1982) 294
\end{thebibliography}
\end{document}